\documentclass[aip,apl,reprint,citeautoscript,floatfix]{revtex4-1}
\usepackage{amsmath,amssymb,amsfonts,bbm,graphicx,color,times}
\usepackage[english]{babel}
\usepackage{float}
\usepackage{color,soul}
\usepackage{siunitx}
\usepackage[percent]{overpic}
\RequirePackage[colorlinks=true, urlcolor=blue, anchorcolor=blue, citecolor=blue, filecolor=blue, linkcolor=blue, menucolor=blue, linktocpage=true, pdfproducer=medialab, pdfa=true]{hyperref} 
\newcommand{\bra}[1]{\ensuremath{\langle #1|}}
\newcommand{\ket}[1]{\ensuremath{|#1 \rangle}}

\newcommand{\ketbra}[2]{\ensuremath{| #1 \rangle\hspace{-2pt} \langle #2 |}}
\newcommand{\braoket}[3]{\ensuremath{\langle #1 | #2 | #3 \rangle}}

\newcommand{\llrr}[1]{\ensuremath{\left( #1\right)}}
\newcommand{\llrrq}[1]{\ensuremath{\left[ #1\right]}}

\renewcommand{\Re}{\operatorname{Re}}

\newcommand{\mathe}{\mathrm{e}}

\begin{document}
\title{All-optical quantum simulator of  qubit noisy channels}
\author{Simone Cialdi}
\email{simone.cialdi@unimi.it}
\homepage{http://users.unimi.it/aqm}
\affiliation{Quantum Technology Lab, Dipartimento di Fisica, 
Universit\`a degli Studi di Milano, I-20133 Milano, Italy}
\author{Matteo A. C. Rossi}
%\email{matteo.rossi@unimi.it}
\homepage{http://users.unimi.it/aqm}
\affiliation{Quantum Technology Lab, Dipartimento di Fisica, 
Universit\`a degli Studi di Milano, I-20133 Milano, Italy}
\author{Claudia Benedetti}
%\email{claudia.benedetti@unimi.it}
\homepage{http://users.unimi.it/aqm}
\affiliation{Quantum Technology Lab, Dipartimento di Fisica, 
Universit\`a degli Studi di Milano, I-20133 Milano, Italy}
\author{Bassano Vacchini}
%\email{bassano.vacchini@unimi.it}
\affiliation{Quantum Technology Lab, Dipartimento di Fisica, 
Universit\`a degli Studi di Milano, I-20133 Milano, Italy}
\affiliation{INFN, Sezione di Milano, I-20133 Milano, Italy}
\author{Dario Tamascelli}
\homepage{http://users.unimi.it/aqm}
%\email{dario.tamascelli@unimi.it}
\affiliation{Quantum Technology Lab, Dipartimento di Fisica, 
Universit\`a degli Studi di Milano, I-20133 Milano, Italy}
\author{Stefano Olivares}
\homepage{http://users.unimi.it/aqm}
%\email{stefano.olivares@unimi.it}
\affiliation{Quantum Technology Lab, Dipartimento di Fisica, 
Universit\`a degli Studi di Milano, I-20133 Milano, Italy}
\author{Matteo G. A. Paris}
%\email{matteo.paris@fisica.unimi.it}
\homepage{http://users.unimi.it/aqm}
\affiliation{Quantum Technology Lab, Dipartimento di Fisica, 
Universit\`a degli Studi di Milano, I-20133 Milano, Italy}
\date{\today}
%%%%%%%%%%%%%%%%%%%%%%%%%%%%%%%%%
\begin{abstract}
We suggest and demonstrate an all-optical quantum simulator for single-qubit noisy channels originating from the interaction with a  fluctuating field. The simulator employs the polarization 
degree of freedom of a single photon, and exploits its spectral components to average over the realizations of the stochastic dynamics. As a proof of principle, we run simulations of dephasing channels driven either by Gaussian (Ornstein-Uhlenbeck) or non-Gaussian (random telegraph) stochastic processes.
\end{abstract}
\maketitle
%%%%%
A quantum simulator (QS) is  a quantum system where the initial preparation 
and  the subsequent time evolution may be controlled and monitored. QSs may
be exploited to mimic the dynamics of other quantum systems that are less
accessible or less controllable \cite{AspuruGuzik2012}. The inherent parallel 
structure of QSs make them suitable to solve problems that are intractable on 
conventional supercomputers, e.g. the simulation of the dynamics of a 
many-particle system. In particular, photonic quantum simulators may be used at room 
temperature, thanks to the fact that photons do not interact with 
each other  \cite{Broome2010,Lanyon2010,Ma2011,Deng2016}. Moreover, photons may propagate in free space 
or in waveguides, and thus may be used to simulate complex structures with 
long range interaction.
\par
In this Letter, we suggest and demonstrate an all-optical QS  
that exploits the spectral components of a single-photon state to perform the parallel 
sum of about one hundred complex numbers. 
In order to demonstrate the operation of our QS, we run the simulation of two different 
single-qubit dephasing channels, arising from the interaction of the quantum system with an external 
fluctuating (stochastic) field. These channels correspond to exact effective models for the interaction 
of qubits with complex environments \cite{Crow2014}, and are found in a variety of physical implementations such as solid-state, superconducting qubits and magnetic systems. In turn, those systems are crucial in the quest for quantum 
technologies, and have been extensively studied  \cite{Xu2013,*LoFranco2012,Paladino2014,Yu2006}. 
Upon exploiting our QS, the interaction with any fluctuating field may be simulated and analyzed.
In particular, here we focus on two paradigmatic channels, driven either by a (Gaussian) Ornstein-Uhlenbeck (OU)
stochastic noise \cite{Benedetti2014b,Rossi2016} or by (non-Gaussian) random telegraph noise (RTN)\cite{Wold2012,Benedetti2014}.
% The understanding of the effects of the interaction
% of a quantum system with the surrounding environment
% is of utmost importance in quantum information processing
% and communication. The unavoidable coupling between system
% and environmental degrees of freedom leads, in fact, to
%losses of coherence and/or energy of the former, thus modifying
% its reduced dynamics. Much attention has been devoted in
% recent years to the possibility that of part of the information
% flown into the environment to come back to the system. When
% present, such back-flow can result, for example, in revivals
% of coherences and are witness of non-Markovian reduced dynamics.
% The reduced dynamics of the system is usually described in
% terms of completely-positive (CP) maps that determine the
% master equation for the density matrix of the system. In the
% case of pure-dephasing maps, the system does not exchange
% energy with the environment; the system/environment interaction,
% therefore, induces only a loss of coherences in the
% system. Depending on the assumptions made on the environment
% and on the system/environment interaction, different approaches
% to the determination of the reduced map are needed.
% In this work we focus on the situation in which the interaction
% with environment can be accounted for by a scalar time dependent
% value extracted form a stochastic process. Being
% this the case, the effective Hamiltonian of the system reads
\par
For a system interacting with a fluctuating field the density operator $\rho(t)$ describing 
the state of the system at any time $t$ corresponds to the average over all possible 
realizations of the stochastic process. The implementation of such a state in the lab
would therefore require the simultaneous generation of a large number
of stochastic trajectories of the process. Here, we show that this procedure 
may be avoided and that the average over the realizations of the noise may be 
obtained in parallel. The quantum information carrier is a photon. The polarization
of the photon is used to encode the state of a qubit, whereas its spectral components
are exploited to implement the trajectories of the stochastic process describing the 
fluctuating field. 
\par
%%%%%%%%%%%%%%%%%%%%%%%%%%%%%%%%%%%%%%%%%%%%%%
We simulate the evolution of a single qubit evolving under a time-dependent Hamiltonian of the form
$H(t) =H_0+H_{\text{int}} = \varepsilon \sigma_z + X(t) \sigma_z$, where $\sigma_z$ is the Pauli matrix
and  $\varepsilon$ determines the energy splitting of the qubit. $H_{\text{int}}$ describes  the 
interaction of the system with a fluctuating environment and 
$X(t)$ is an arbitrary real-valued continuous-time stochastic process. 
The environment induces decoherence, but does not exchange energy with the system.
If the qubit is initially prepared in the state $\ket{\psi_0} = ({1}/{\sqrt{2}})\left( \ket{0} + \ket{1}\right) $, 
the evolved state is given by $\rho(t) = \left \langle U(t)\rho_0 U^\dagger(t) \right \rangle$, where
$\rho_0 = \ket{\psi_0}\bra{\psi_0}$, $U(t) = \exp[-i\int_0^t H(\tau)d\tau]$ is the evolution operator 
and $\langle \cdot \rangle$ denotes the expectation value over the realizations of the stochastic process, i.e. 
of the noise.  In the interaction picture we have
\begin{equation}
    \rho(t) %= V(t) \rho_0 
= \frac{1}{2}
    \llrr{ 
    \begin{array}{c c}
        1 &  \left\langle e^{-2i  \Phi(t)}\right\rangle  \\
    \left\langle e^{2 i \Phi(t)}\right\rangle  & 1
    \end{array}
},
       \label{eq:meanEvolvedState}
\end{equation}
where $\Phi(t) = \int_0^t X(\tau)d\tau$.
%where $V(t)$ is the dynamical map determined by the process $X(t)$.
%
%\section{Quantum simulation}
In order to obtain the state of the system at any time $t$ we should
compute the average of a sufficiently large collection of independent
realizations (sample-paths) of the stochastic process
 $X(t)$. 
 Each sample-path is a real scalar function $\Phi_r(t) = \int_0^t X_r(\tau)d\tau$, that 
 corresponds to the phase shift induced by a particular realization $X_r(\tau)$, with $r$ running on 
the sample index. 
\par
\begin{figure}[t]
\includegraphics[width=.95\columnwidth]{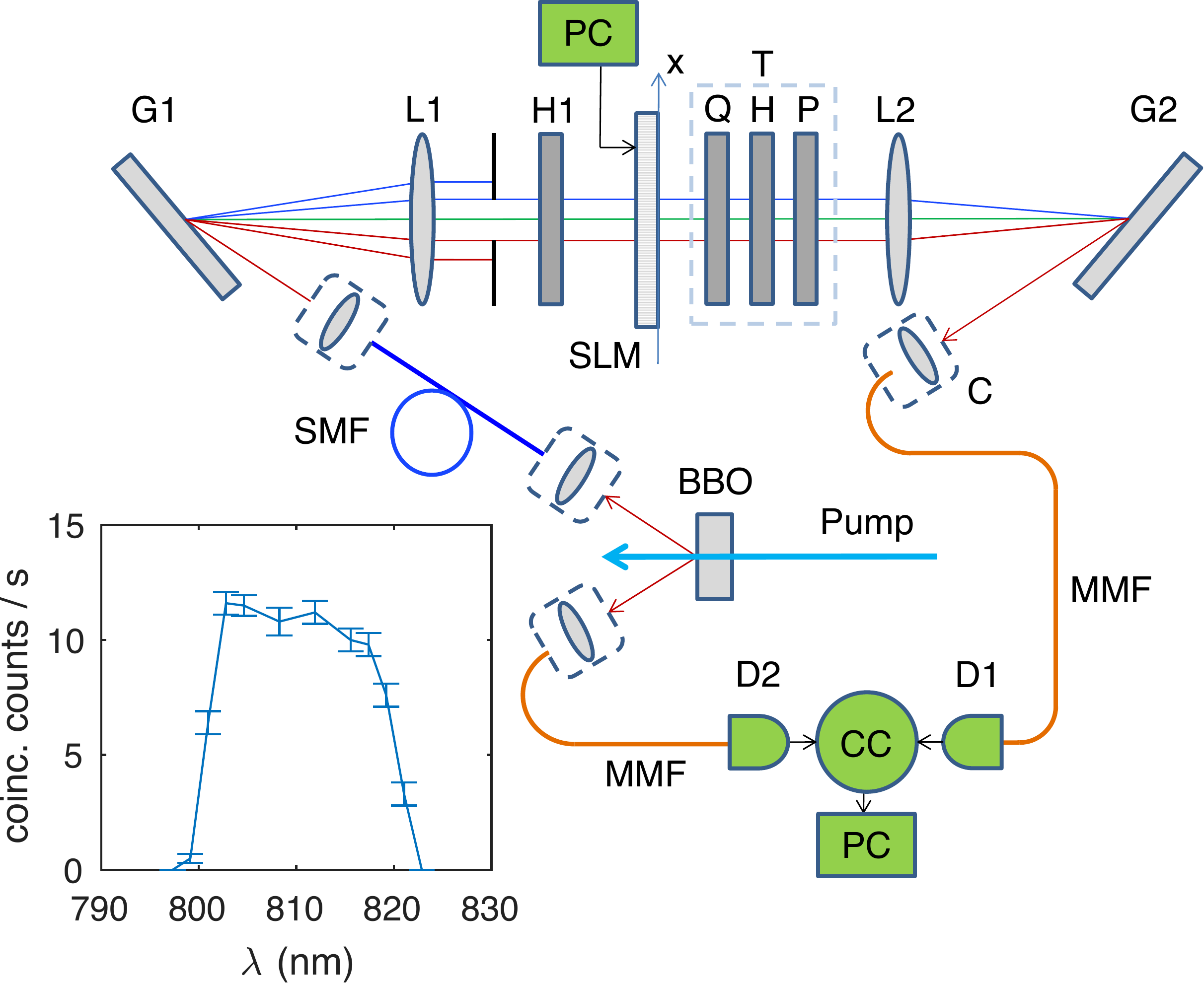}
\caption{Schematic diagram of our 
setup. Pump, 405.5 nm laser diode; BBO, Beta barium borate nonlinear crystal; SMF, single-spatial-mode
and polarization preserving fiber; MMF, multimode fiber; G1-G2, gratings; L1-L2, lens; H1, half-wave-plate;
SLM, spatial light modulator; T, tomographic apparatus; C, optical coupler; D1-D2, single photon detectors;
CC, coincidences counter. Inset shows the measured PDC spectrum.}
\label{fig:apparatus_spectrum}
\end{figure}

In the following we describe an experimental all-optical setup that
allows us to obtain the evolved state upon the generation of $n$
sample-paths in a single run. In particular, the qubit (polarization)
state at time $\bar{t}$ will be given by $\rho(\bar{t})= \frac{1}{n}
\sum_{r=1}^n \ketbra{\psi_r(\bar{t})}{\psi_r(\bar{t})}$, where,
according to Eq.~\eqref{eq:meanEvolvedState}, $\ket{\psi_r(\bar{t})} =
({1}/{\sqrt{2}}) \left( e^{-2i \Phi_r(\bar{t})} \ket{H}+\ket{V}\right)
$.  In Fig.~\ref{fig:apparatus_spectrum} we show a schematic diagram of
the experimental apparatus. The frequency-entangled two-photon state is
generated by parametric down-conversion (PDC) with a diode pump laser @
\SI{405.5}{\nano\meter} by using a BBO crystal (\SI{1}{\milli\meter}
thick).  The laser is temperature stabilized and generates
\SI{40}{\milli\watt} @ \SI{70}{\milli\ampere}.  The two photons are then
collected by two fiber couplers and sent respectively into a
single-spatial-mode and polarization-preserving fiber (SMF) and a
multimode fiber (MMF).  When the idler photon enters the coupler, it
travels entirely through the fiber towards the single photon detector
(D2). Conversely, the signal photon, after a short fiber (SMF), enters a
4F system \cite{Weiner2000}, i.e. propagates in the air, through few
optical devices [the gratings G1 and G2
(\SI{1714}{\text{lines}/\milli\meter}) 
and lens L1 and L2 ($f= \SI{500}{\milli\meter}$)] an half-wave plate (H1),
that we use for the input state preparation, a spatial light modulator (SLM)
and a tomographic apparatus (T) \cite{Banaszek1999,James2001} to reconstruct the output state.
At the end of the 4F system the signal photon is coupled to a multimode fiber
and reaches the single photon detector (D1).
Finally an electronic device measures the coincidence counts (CC) and sends them
to the computer (PC).
The tomographic apparatus (T) is composed of a quarter-wave plate (Q), an 
half-wave plate (H) and a polarizer (P). 
The SLM is a 1D liquid crystal mask (640
pixels, \SI{100}{\micro\meter/pixel}) and is placed on the Fourier plane between the two
lenses L1 and L2 of the 4F system (see Fig.~\ref{fig:apparatus_spectrum}). 
The SLM is controlled by the computer (PC) and is used
to introduce a different phase $\Phi_{r}(\bar{t})$ for each pixel.
In the Fourier plane the spectral components of the signal photon are linearly 
dispersed (\SI{1.82}{\nano\meter/ \milli\meter}).
%%%
\par
%%%
In order to measure the PDC spectra we used a \SI{2}{\milli\meter} slit on
the Fourier plane of the 4F system.  We calibrated the slit using a
graduated reference on the Fourier plane and for each slit position
(and therefore for each wavelength) we recorded coincidence counts
from the detectors. In the inset of Fig.~\ref{fig:apparatus_spectrum} we show the measured PDC
spectrum.  We observe that it is selected by the limited
width of the H1 plate mount, in such a way that the intensity of the
spectral components impinging on the SLM is almost constant, a relevant 
feature to implement our QS. For this reason we are limited to
use $n=100$ out of the $640$ pixel available on the SLM.
\par
When leaving the BBO, signal ($s$) and idler ($i$) photons are in the pure state
$\int d\omega f(\omega) \ket{H}_{s} \otimes \ket{\omega}_{s} \otimes  \ket{H}_{i} \otimes \ket{-\omega}_{i} $ 
\cite{Joobeur1996},
where $H$ denotes the horizontal polarization and $\omega$ is the
spectral shift with respect to the PDC central component
$\omega_0 = \omega_p/2$ where $\omega_p$ is the pump laser
frequency. We point out that the polarization and frequency degrees of
freedom of the two photons are independent of each other and thus,
upon the detection of an idler photon, the conditional state of the signal photon, i.e. 
the partial trace over the idler degrees of freedom, is given by the mixed state
\begin{equation}
    \rho_{SE} = \rho_S \otimes \rho_E = \ketbra{H}{H} \otimes \int d\omega |f(\omega)|^2 \ketbra{\omega}{\omega}. 
    \label{eq:afterTrace}
\end{equation}
The initial system-environment state is therefore factorized, and this warrants the existence of 
the reduced dynamics \cite{Pechukas1994}.
%%%
\begin{figure*}[t]
\includegraphics{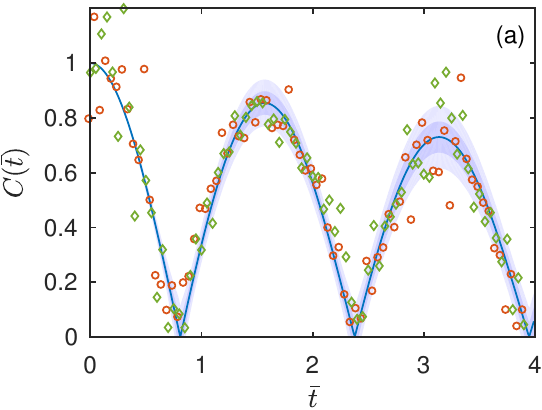}
\includegraphics{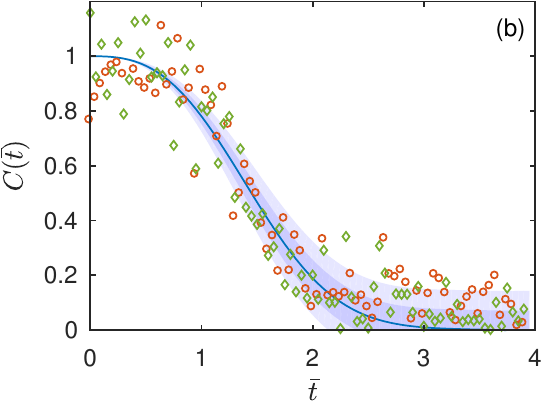}
\includegraphics{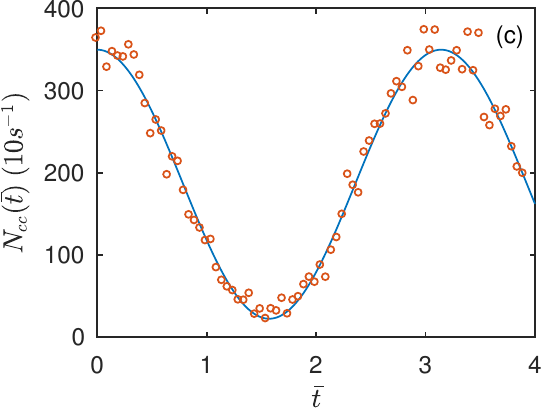}
\caption{\label{fig:tomo} (a) and (b): Dynamics of the
off-diagonal element of $\rho_S(\bar t)$, $C(t)= |\langle
    e^{-2i\Phi_r(\bar{t})}\rangle_n|$, for RTN (a) and OU (b) with $\gamma = 0.1$. Red circles and green diamonds represent the data obtained, respectively, with tomographic reconstruction of $\rho_{S,\text{exp}}(\bar t)$ and projection onto the state $\ket{+}$. The blue line is the analytic solution of the model. The shades represent intervals of $1\sigma$ (darker) and $2\sigma$ (lighter) around the analytic solution, where $\sigma$ is the standard deviation of paths obtained with 100 realizations of the stochastic process. Note that the noise for small $\bar t$ is due to the Poissonian fluctuations on the coincidence counts.
    (c): Coincidence counts $N_{\text{cc}}(\bar{t})$ in the case
 of RTN with $\gamma=0$, the blue line is the fit with the function
$N_\text{cc}=N(1+p \: \cos(2 \bar{t}))$.
}
\end{figure*}
%%%
The polarization of the idler photon encodes a qubit, while the spectral/spatial degrees of freedom may 
be considered as the environment. The grating G1 (see Fig.~\ref{fig:apparatus_spectrum}) disperses 
linearly the photon spectral components $\omega$ and the lens L1 focuses them on the Fourier plane 
of the 4F system where the SLM is placed. Each spectral component $\omega$ is characterized by a
Gaussian spatial profile (\SI{60}{\micro\meter} FWHM) centered in the spatial coordinate $x$. We 
have $\omega=\alpha x$ where $\alpha=\SI{1.82}{\nano\meter/\milli\meter}$. In order to emphasize 
that the spectral components are spatially dispersed we use the notation $\ket{x} = \ket{\omega(x)}$. 
The half-wave plate H1 rotates the polarization of the signal photon, turning the initial state of the 
system to $\ket{\psi_S(0)} = ({1}/{\sqrt{2}})  \left( \ket{H} + \ket{V}\right) $.
Introducing with the notation \ket{\eta_r}  the $r$-th pixel of the SLM we have
$\ket{x} = \sum_r \eta_r(x) \ket{\eta_r}$, where $|\eta_r(x)|^2$
is the probability that the component $x$ passes through the $r$-th
pixel. In this notation the identity $\mathbbm{1} = \sum_r \ketbra{\eta_r}{\eta_r}$ expresses the 
fact that all detectable components pass through the pixels.
\par
The initial state of the environment then reads 
$\rho_E= \sum_{r,s} A_{rs} \ketbra{\eta_r}{\eta_s}$, where
\begin{equation}
    A_{rs}=\int dx |f(x)|^2 \eta_r(x) \eta_s^*(x).
    \label{eq:Ars}
\end{equation}
The matrix $A_{rs}$ is positive definite with trace equal to one. 
The SLM imprints a pixel-dependent phase on the horizontal polarization component, 
which we denote by $e^{ -2 i \Phi_r(\bar{t})}$ (see Eq.~\eqref{eq:meanEvolvedState}). The unitary interaction operator can therefore be written in the form
\begin{align}
    U(\bar{t}) = \exp \llrrq{-2 i P_H \otimes \sum_r \Phi_r(\bar{t}) P_r},
    \label{eq:propagator}
\end{align}
where $P_H = \ketbra{H}{H}$ and $P_r = \ketbra{\eta_r}{\eta_r}$. As a result
$U(\bar{t}) \ket{H} \otimes \ket{\eta_r} = e^{-2i \Phi_r(\bar{t})}
\ket{H} \otimes \ket{\eta_r}$, while vertically polarized states are
left unchanged.
%  and 
% $U(\bar{t}) \ket{V} \otimes \ket{\eta_r} = \ket{V} \otimes \ket{\eta_r}$.
Taking the marginal of  $\rho_{SE} ( \bar{t} )  =  U ( \bar{t} ) \llrr{\rho_{S} ( 0 ) \otimes \rho_{E}} U ( \bar{t} )^{\dag}$
we thus obtain
%\begin{widetext}
\begin{equation}
  \rho_{S} ( \bar{t} ) =  \frac{1}{2}\sum_{r} A_{rr}   
   \left(\begin{array}{cc}
    1 & \mathe^{-2i \Phi_{r} ( \bar{t} )}\\
    \mathe^{2 i \Phi_{r} ( \bar{t} )} & 1
  \end{array}\right),
\end{equation}
so that the only matrix element affected by the dynamics is \braoket{H}{\rho_{S} ( \bar{t} )}{V}.
In our case for the diagonal elements we have $A_{rr}={1}/{n}$ ($n=100$)
because the selected PDC  spectrum is basically rectangular. However, due to the 
imperfections of the experimental apparatus, in each realization the state is not exactly pure but rather of
the form $\rho_{S,\text{exp}} = p \rho_{S} + (1-p)\rho_{\text{mix}}$, where 
$\rho_{\text{mix}} =\frac{1}{2} \ket{H}\bra{H} + \frac{1}{2}
\ket{V}\bra{V}$ is the maximally mixed state, so that the relevant
quantity to be measured is  
\begin{equation}
  \label{eq:1}
  \braoket{H}{\rho_{S,\text{exp}} ( \bar{t} )}{V}=\frac12\, p \left\langle e^{-2i  \Phi_r(\bar{t})}\right\rangle_n.
\end{equation}
% Finally, we obtain: 
% \begin{equation}
%     \rho_{S,\text{exp}}(\bar{t}) = \frac{1}{2}
%     \llrr{ 
%     \begin{array}{c c}
%         1 &  p \left\langle e^{-2i  \Phi_r(\bar{t})}\right\rangle_n  \\
%     p \left\langle e^{2 i \Phi_r(\bar{t})}\right\rangle_n  & 1
%     \end{array}}.
%        \label{eq:meanr}
% \end{equation}
In our setup, the average over the realizations of the noise is
performed by (coherently) 
collecting the different spatial components $\ket{\omega}$
through the lens L2 and the grating G2 into a multimode fiber. The state
reconstruction is performed by the tomographic apparatus T placed
between the SLM and the lens L2.  
\par
In the followin, we show the results obtained by running simulations of
two dephasing channels driven either by Ornstein-Uhlenbeck Gaussian
noise or non-Gaussian random telegraph noise.  Apart from providing a
convenient description of many realistic environments, dephasing
channels also permit a simple assessment of the non-Markovian character
of the reduced dynamics of the system \cite{Breuer2009}.  This criterion
relies on the study of the behaviour in time of the distinguishability
among different initial states of the system evolved according to the
same reduced dynamics. The distinguishability between states is
quantified by their trace distance defined as $D(t) = \frac 12 \|
\rho_1(t)-\rho_2(t) \|_1$, that is half the trace norm of the difference
of the two statistical operators.  Non-Markovianity is associated to
revivals in time of this quantity.  In particular, it can be
shown\cite{Breuer2016a} that for a dephasing map the highest sensitivity
to these revivals is obtained by looking at the modulus of the
coherences of the statistical operator $\rho(t)$ of
Eq.~\eqref{eq:meanEvolvedState}, which indeed equals the trace distance
among the pair of states better witnessing non-Markovianity.
\par
For the RTN, the realization $X_r(\bar{t})$ flips randomly between the
values $\pm 1$ with a switching rate $\gamma$. In our case for each step
of the realization the simulation time $\bar{t}$ is incremented by
$\delta \bar{t} = 0.001$ in units of $1/ \gamma$.  The flip probability
at each step is given by $\delta P= 1-e^{-\gamma \delta \bar{t}}$.  The
initial values $X_r(0)$ are selected randomly with equal probability
between $\pm1$ for each pixel. In the case of the OU process we have:
\begin{equation}
    X_r(\bar{t} + \delta \bar{t} ) = \left( 1-2\gamma \delta \bar{t} \right)  
    X_r(\bar{t}) + 2 \sqrt{\gamma} \: dW(\bar{t}),
    \label{eq:ou-sde}
\end{equation}
where $dW(\bar{t})$ is a Wiener increment with mean equal to zero and
standard deviation $\sigma = \sqrt{\delta \bar{t}}$. For each
realization (i.e. for each pixel) we impose the initial condition
$X_r(0)=0$.  Both models are analytically
solvable\cite{Benedetti2014,Rossi2014}, and it is known that any
dephasing map induced by a Gaussian stochastic process is Markovian,
while RTN gives a non-Markovian map for $\gamma <
2$.\cite{Benedetti2014}
In Fig.~\ref{fig:tomo}(a) and Fig.~\ref{fig:tomo}(b) we plot the
experimental results in the case of the RTN and OU process respectively.
In both cases we have $\gamma = 0.1$ in arbitrary units.  We note the
presence of strong revivals in the RTN case, according to the
non-Markovian character of the dynamics. In the OU case the off-diagonal
element of $\rho_S(\bar t)$ decays monotonically, as expected for a
Markovian dynamics. For each point of the graph ($\bar{t}_i=i \times 50
\delta \bar{t} $) we send to the pixels the phases $\Phi_r(\bar{t}_i) =
\int_0^{\bar{t}_i} X_r(\tau)d\tau$ and we reconstruct the state with the
tomographic method by performing four projective
measurements\cite{Banaszek1999,James2001,Cialdi2014}. We use an
acquisition time of $\SI{10}{\second}$ for each measure of coincidence
counts.  For a pure
dephasing dynamics one has:
\begin{equation}
    D(t) = |\langle e^{-2i  \Phi(t)}\rangle | 
    \approx |\langle e^{-2i  \Phi_r(\bar{t})}\rangle_n | \equiv C(t)\,.
\end{equation}
Notice that in order to obtain the non-Markovianity from the revivals of
the trace distance we need the factor $\frac12 p$. Indeed, while the
trace distance is in principle  bounded by one, here we estimate its
value from the reduced dynamics of the off-diagonal  matrix elements,
whose actual value depends on the purity of the system state. The latter
is known only in average and it is also affected by experimental
uncertainty due to the Poissonian statistics of photon counting. The
quantity $C(t)$  is shown in Fig.~\ref{fig:tomo}(a) and
Fig.~\ref{fig:tomo}(b) as a function of $\overline t$ for RTN and OU
noise, both with $\gamma = 0.1$. Notice that $\langle e^{-2i
\Phi(\bar{t})}\rangle$ is real-valued because the two considered
stochastic processes have zero mean (and indeed, from the tomographic
measures, we find that the imaginary part of $\langle e^{-2i
\Phi_r(\bar{t})}\rangle_n$ is zero within the experimental uncertainty).
Thus, in order to estimate the trace distance we can perform just one
projective measure on the state $\ket{+} =
({1}/{\sqrt{2}}) \left( \ket{H} + \ket{V}\right)$, since we have
$\bra{+} \rho_{S,\text{exp}} \ket{+} = \frac{1}{2} \left( 1+ p 
\Re \left\langle e^{-2i  \Phi_r(\bar{t})}\right\rangle_n    \right) $.
In order to obtain the parameter $p$ we acquire 
a reference measure using the RTN with $\gamma = 0$ (i.e., static noise). In this case 
we have $\langle e^{-2i  \Phi_r(\bar{t})}\rangle = \cos(2\bar{t} )$.
In Fig.~\ref{fig:tomo}(c) we can see the coincidence counts vs. the simulation time 
$\bar{t}$ in the case of the RTN with $\gamma=0$.
From the fit (blue solid line) with the function
$N_\text{cc}(\bar{t})=N(1+p \: \cos(2\bar{t}))$ we find $p=0.88 \pm
0.02$ as well as $N=186 \pm 2$. Thus, in the general case we can write: $\left\langle e^{-2i  \Phi_r(\bar{t})}\right\rangle_n
= ({N_\text{cc}(\bar{t})-N})/{p}$.
In Fig.~\ref{fig:tomo}(a) and \ref{fig:tomo}(b) we can also see the
comparison between the tomographic method (red circles) and the method
based on the projection on the state $\ket{+}$ (green diamonds) in the
case of the RTN and of the OU. We note that the two methods indeed give compatible results.
\begin{figure}[t]
\includegraphics{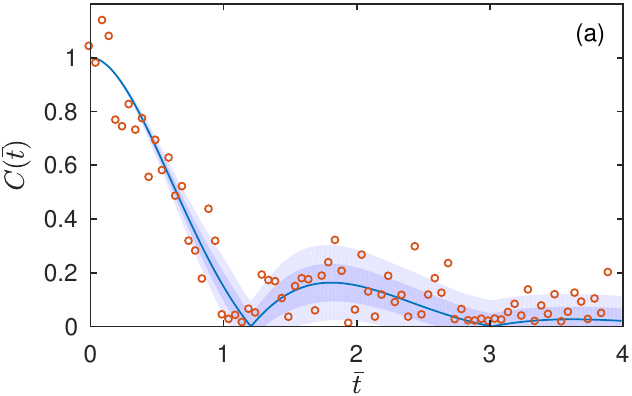}
\vspace{3pt}
\includegraphics{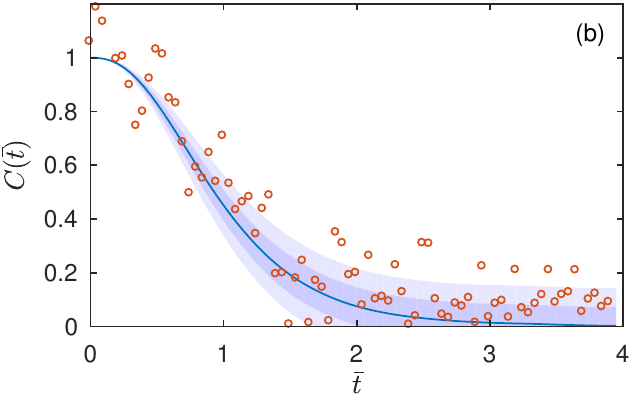}
\caption{\label{fig:simple} Dynamics of the off-diagonal element of $\rho_S(\bar t)$, $C(t)= |\langle
    e^{-2i\Phi_r(\bar{t})}\rangle_n|$, vs $\bar{t}$ evaluated by the method of the projection onto the state
$\ket{+}$ in the case $\gamma=1$  for RTN (a) and OU (b) stochastic process. The blue line is the analytic solution and the blue shades represent intervals of $1\sigma$ (darker) and $2\sigma$ (lighter) around the analytical solution, where $\sigma$ is the standard deviation of paths obtained with 100 realizations of the stochastic process.}
\end{figure}
In Fig.~\ref{fig:simple} we can see the results obtained by the projection method on the state
$\ket{+}$ and with $\gamma=1$, for both RTN (a)
and OU process (b). Note the decrease of non-Markovianity of the RTN dynamics compared 
 to the case with $\gamma=0.1$. In turn, the non-Markovianity vanishes when $\gamma 
\ge 2$ \cite{Benedetti2014}. In the case of the OU process the dynamics remains Markovian 
as expected.
\par
In conclusion, we have suggested and demonstrated an all-optical 
quantum simulator for single-qubit noisy channels. 
The simulated qubit is encoded in the polarization degree of freedom of a 
single-photon generated by parametric downconversion, whereas
several realizations of the noise are achieved in a single shot 
by using a programmable spatial light modulator on 
the different spectral components of the photon.
\par
As a proof of principle, we have run simulations of dephasing channels driven either by Gaussian 
(Ornstein-Uhlenbeck) or non-Gaussian (random telegraph) stochastic processes.
Upon increasing the number of pixels in the spatial light modulator one may increase the number of realizations
and perform more accurate simulations of noisy channels and complex classical environments.
\\
\begin{acknowledgments}
This work has been supported by EU through the collaborative project
QuProCS (Grant Agreement 641277) and by UniMI through the H2020 
Transition Grant.
\end{acknowledgments}
\bibliography{aoqsqnc}

%%%%%%%%%%%%%%%%%%%

\end{document}